 \documentclass[final,5p,times,twocolumn]{elsarticle}

\usepackage{amssymb}
\usepackage{lipsum}
\usepackage{graphicx} % Required for inserting images
\usepackage{amsmath,esint,mathtools}
\usepackage{physics}
\usepackage{graphicx}
\usepackage{xcolor}
\usepackage{amssymb,amsthm}
\usepackage{enumerate} 
\usepackage{tikz}
\usepackage{cancel}
\journal{Nuclear Physics B}

\begin{document}

\begin{frontmatter}



\title{Integral Transformations for Conformally Invariant Celestial Amplitudes} %แก้ชื่อหลังจากเพิ่มกราวิตอน


\author[first]{Aphiwat Yuenyong}
\ead{yuenyong.aphiwat47@gmail.com}
\author[second]{Pongwit Srisangyingcharoen\corref{cor1}}
\ead{pongwits@nu.ac.th}
\author[first]{Ekapong Hirunsirisawat}
\ead{ekapong.hir@kmutt.ac.th}
\author[first]{Tanapat Deesuwan}
\ead{tanapat.dee@kmutt.ac.th}
\affiliation[first]{organization={Quantum Computing and Information Research Centre (QX), Department of Physics, Faculty of Science, King Mongkut's University of Technology Thonburi},%Department and Organization
            %addressline={}, 
            %city={},
            %postcode={}, 
            state={Bangkok},
            country={Thailand}}
\affiliation[second]{organization={The Institute for Fundamental Study, Naresuan University},%Department and Organization
            %addressline={}, 
            %city={Earth},
            postcode={65000}, 
            state={Phitsanulok},
            country={Thailand}}
\cortext[cor1]{Corresponding author}
\begin{abstract}
We propose an integral transformation for celestial amplitudes for massless particles that maps the celestial coordinates \((z_i,\bar z_i)\) to a new set of complex variables \((s_i,\bar s_i)\), inspired by the structure of closed string scattering amplitudes. A consistent inverse transformation is constructed by regulating a divergence associated with translational redundancy and absorbing it into an overall normalization. Applying this transformation to celestial MHV amplitudes, we derive constraints on \((s_i,\bar s_i)\) for three-, four-, and general \(n\)-point amplitudes, and show that these conditions are necessary for invariance under global conformal transformations. The physical data of the external particles are encoded in the new variables $(s_j,\bar{s}_j)$.
\end{abstract} %แก้เพิ่ม




\end{frontmatter}

%\tableofcontents



\section{Introduction}\label{intro}
Celestial holography proposes a duality between four-dimensional asymptotically
flat spacetimes and a two-dimensional conformal field theory living on the
celestial sphere at null infinity \cite{Strominger:2017zoo,
raclariu2021lecturescelestialholography,deBoer:2003vf}. In this framework, four-dimensional
scattering amplitudes can be mapped onto the celestial sphere via the Mellin
transformation over the external particle energies, which turns momentum eigenstates into boost eigenstates. The resulting celestial amplitudes behave as conformal
correlation functions and transform covariantly under global conformal
transformations \cite{Pasterski:2016qvg,Pasterski:2017ylz,Pasterski:2017kqt}. A central feature of celestial holography is that soft theorems of the bulk theory are reinterpreted as Ward identities of the dual conformal field theory associated with asymptotic symmetries at null infinity \cite{Strominger:2013jfa,Strominger:2013lka}.

Scattering amplitudes play a central role in quantum field theory, serving as a
bridge between theoretical predictions and experimental observations. Within the celestial framework, tree-level
celestial gluon amplitudes were first derived in \cite{Pasterski:2017ylz} for
low-point cases. Subsequently, it was generalized to arbitrary \(n\)-point in \cite{SCHREIBER2018349}. Furthermore, tree-level string
scattering amplitudes can be mapped onto the celestial sphere, as shown in
\cite{StringOnCS}. At the one-loop level, celestial open- and closed-string
amplitudes were formulated in
\cite{Donnay:2023kvm,CanazasGaray:2025xlh}. Many attempts have been made to reveal the structural properties of celestial amplitudes, including the on-shell recursion relations \cite{Pasterski:2017ylz}, double copy structures
\cite{Casali:2020vuy}, and the monodromy relations \cite{Adamo:2024mqn}. 

String scattering amplitudes exhibit remarkable structural relations as a consequence of conformal invariance on the worldsheet. Notably, the Kawai--Lewellen--Tye (KLT) relations allow closed-string amplitudes to be factorized into products of open-string amplitudes \cite{Kawai:1985xq}. In the low-energy limit, these relations reduce to the celebrated double-copy relations connecting gauge theory and gravity amplitudes \cite{Bern:2002kj,Bern:2008qj,Bern:2010ue,Bern:2010yg}.

The worldsheet integrals appearing in string amplitudes play a central role in establishing such structural relations. Examples include monodromy relations among color-ordered open-string amplitudes \cite{bjerrum2009minimal}, relations expressing mixed open--closed string amplitudes in terms of purely open-string ones \cite{Stieberger:2009hq,Stieberger:2015vya}, connections between closed-string and mixed-string amplitudes \cite{Yuenyong:2024ebe}, and string on-shell recursion relations \cite{Boels:2010bv,Chang:2012qs,Srisangyingcharoen:2024qyx,Srisangyingcharoen:2024zko}.

A notable feature of the standard celestial construction is that the Mellin transformation acts only on the energy variables, while the momentum variables remain unchanged. As a result, energies and momentum directions are treated asymmetrically, and celestial amplitudes transform only covariantly, rather than invariantly, under conformal transformations. This motivates the search for alternative representations in which the four-dimensional kinematical data are encoded more uniformly. In this work, we investigate a new class of Mellin-like integral transformations acting on momentum variables in addition to energies, leading to a new set of transformed variables whose relation to the external particle data is discussed in Section~\ref{comment}.

A further motivation comes from string theory, where exact worldsheet conformal invariance underlies powerful structures such as monodromy relations and worldsheet factorizations. Since these techniques cannot be directly implemented in the standard celestial basis, a conformally invariant formulation may provide a more natural framework for importing such methods into celestial holography and for uncovering new structural properties of this holography.

Motivated by these observations, we introduce a class of integral transformations inspired by the structure of worldsheet integrals in string theory and investigate their role in constructing conformally invariant celestial amplitudes.

The organization of the paper is as follows. In Section~\ref{prelim}, we review scattering amplitudes and celestial amplitudes, and present the three- and four-point celestial gluon and graviton amplitudes. Section~\ref{construct} is devoted to the construction of the integral transformation and its inverse, beginning with the three- and four-point cases and subsequently generalizing to the 
$n$-point case. In Section~\ref{TestFor}, we demonstrate that, under the integral transformation introduced in Section~\ref{construct}, the transformed celestial amplitudes naturally emerge the conditions required for conformal invariance. Explicit calculations are presented for the three-, four-, and $n$-point cases. In Section \ref{comment}, we show the explicit relations between physical data of external particle $(\Delta_j,J_j)$ and the new variables $(s_j,\bar{s}_j)$. Finally, we summarize our findings and discuss future directions in Section~\ref{conclusion}. %แก้เพิ่่มsection

\section{Preliminaries}\label{prelim}
\subsection{Scattering amplitudes}

A general form of an \(n\)-point scattering amplitude is given by
\begin{align}
    A_{l_1 l_2 \dots l_n}(k_j^{\mu})
    =  \delta^{(4)}\!\left(\sum_{j=1}^{n} k_j^{\mu}\right)
    A^{\text{YM}}_{l_1 l_2 \dots l_n} \,
\end{align}
where $l_i$ labels helicities of external particles. The four-momentum of a massless external particle can be parametrized as
\begin{align}
    k_j^{\mu} &= \epsilon_j \omega_j q_j^{\mu}\nonumber \\
    &= \frac{\epsilon_j \omega_j}{2} \bigl(
    1+|z_j|^2,\,
    z_j+\bar{z}_j,\,
    -i(z_j-\bar{z}_j),\,
    1-|z_j|^2
    \bigr),
\end{align}
where $z_j$ and $\bar{z}_j$ are the complex coordinates on the celestial sphere, $q^{\mu}_j$ is a  null vector pointing toward the celestial sphere, $\omega_j$ is the angular frequency associated with the energy of the external particle, and \(\epsilon_j = \pm 1\) denotes outgoing and incoming particles, respectively.  

In this paper, we focus on the maximal helicity violating (MHV) gluon scattering amplitudes in pure Yang-Mills (YM) theory. The corresponding $n$-point amplitudes \(A_{l_1 l_2 \dots l_n}^{\text{YM}}\) are given by the Parke--Taylor formula \cite{ParkeTaylor},
\begin{align}
    A_{1^+ 2^+ \dots i^- \dots j^- \dots n^+}^{\text{YM}}
    =
    \frac{\langle ij \rangle^4}
    {\langle 12 \rangle \langle 23 \rangle \langle 34 \rangle \cdots \langle n1 \rangle},
\end{align}
where \(\langle ij \rangle\) denotes the spinor inner product. The spinor products can be expressed in terms of the complex coordinates
and energies as
\begin{align}
    \langle ij \rangle = \sqrt{\omega_i \omega_j}\, z_{ij},
    \qquad
    [ij] = - \sqrt{\omega_i \omega_j}\, \bar{z}_{ij},
\end{align}
with \(z_{ij} \equiv z_i - z_j\) and
\(\bar{z}_{ij} \equiv \bar{z}_i - \bar{z}_j\). The Mandelstam variables read
\begin{equation}
    s_{ij}=2k_i\cdot k_j=\langle i j\rangle[ji]=\omega_i\omega_j z_{ij}\bar{z}_{ij}.
\end{equation}

According to the double-copy relations \cite{Bern:2002kj,Bern:2008qj,Bern:2010ue,Bern:2010yg}, the MHV graviton amplitude can be expressed in terms of the square of Yang-Mills amplitudes. The $n$-point graviton scattering amplitudes  $A^{G}_{l_1l_2\dots l_n}$ are given by \cite{Elvang_Huang_2015}
\begin{align}
    A^{G}_{1^-,2^-,3^+,\dots ,n^+}  = \sum_{P(i_3,i_4,\dots,i_n)}s_{1i_n}\left(\prod_{k=1}^{n-1} \beta_k \right) \left(A^{\text{YM}}_{1^-,2^-,i^+_3,i^+_4,\dots,i^+_n}\right)^2
\end{align}
where 
\begin{align}
     \beta_k &= - \frac{\langle i_k, i_{k+1} \rangle}{\langle 2,i_{k+1}\rangle}\langle 2| i_3+i_4+\dots+i_{k-1}|i_k] \nonumber \\
     &= - \frac{\langle i_k, i_{k+1} \rangle}{\langle 2,i_{k+1}\rangle} \sum_{m=3}^{k-1}\langle 2, i_m\rangle [i_m, i_k].
\end{align}
%Physically, the double-copy relation encodes a correspondence in which graviton amplitudes arise as a square of gauge-theory amplitudes, highlighting a deep structural link between gravity and Yang--Mills theory. 

\subsection{Celestial amplitudes}
The celestial amplitude for massless particles is obtained by applying a Mellin
transformation to momentum-space scattering amplitudes
\cite{Pasterski:2017ylz,Pasterski:2017kqt}. The \(n\)-point celestial amplitude for
massless particles is defined as
\begin{align}
    \mathcal{A}_n(\lambda_j,z_j,\bar{z}_j)= \int_{0}^{\infty} \prod_{j=1}^{n}d\omega_j \, \omega_j^{\Delta_j-1}A_n(\omega_j,z_j,\bar{z}_j)
\end{align}
where $\Delta_j$ is conformal dimension, which is equal to $1+i\lambda_j$, $\lambda_j\in\mathbb{R}$.  The Mellin transformation maps plane-wave solutions in energy space to conformal primary wavefunctions on the celestial sphere.
Accordingly, the celestial amplitude
\(\mathcal{A}_n(\lambda_j, z_j, \bar{z}_j)\) exhibits the structure of a conformal
correlation function on the celestial sphere. Under global conformal transformation,
\begin{align}
    z_j \to z'_j = \frac{az_j+b}{cz_j+d} \qquad \text{and} \qquad \bar{z}_j \to \bar{z}_j = \frac{\bar{a} \bar{z}_j + \bar{b}}{\bar{c} \bar{z}_j + \bar{d}}, \label{globalcon}
\end{align}
where $a,b,c,d \in \mathbb{C}$ and $ad-bc=1$, the celestial amplitude transforms covariantly as 
\begin{align}
     \mathcal{A}_{J_1 \dots J_n}\!\left(\lambda_j, 
    \frac{a z_j + b}{c z_j + d},
    \frac{\bar{a} \bar{z}_j + \bar{b}}{\bar{c} \bar{z}_j + \bar{d}}\right)
    = \prod_{j=1}^{n} 
   (c z_j + d)^{\Delta_j + J_j} 
    \nonumber \\
    \times (\bar{c} \bar{z}_j + \bar{d})^{\Delta_j - J_j}  \mathcal{A}_{J_1 \dots J_n}(\lambda_j, z_j, \bar{z}_j).
\end{align}

In four-dimensional Minkowski spacetime with signature $(+---)$, three-point
scattering amplitudes of massless particles vanish as a consequence of momentum
conservation, since the on-shell conditions force all momenta to be collinear.
However, upon analytically continuing to the $(++--)$ signature, one may treat
the complex coordinates $z_j$ and $\bar{z}_j$ as two linearly independent
variables. This allows for nontrivial kinematics, and in particular the MHV
three-point gluon amplitudes need not vanish. According to the Parke--Taylor
formula, the nonvanishing three-point amplitude takes the form
\begin{align}
    A_{--+} = \delta^{(4)}(\omega_1q_1+\omega_2q_2-\omega_3q_3)\frac{\omega_1\omega_2}{\omega_3}\frac{z_{12}^{3}}{z_{13}z_{23}}.
\end{align}
The momentum conservation in the delta function can be rewritten as 
\begin{align}
    \delta^{(4)}(\omega_1q_1+\omega_2q_2-\omega_3q_3) = &\frac{4}{\omega_3^2z_{23}z_{31}}\delta(\omega_1-\frac{z_{32}}{z_{12}}\omega_3) \nonumber\\
    &\times \delta (\omega_2 - \frac{z_{31}}{z_{21}}\omega_3)\delta(\bar{z}_{13})\delta(\bar{z}_{23}).
\end{align}
Applying the Mellin transformation, it was shown in \cite{StringOnCS} that the three-point celestial gluon amplitude is given by
\begin{align}
    \mathcal{A}_{--+} =8\pi\delta(\bar{z}_{13})\delta(\bar{z}_{23})\delta(\sum_{j=1}^{3}\lambda_j) z_{21}^{1-i(\lambda_1+\lambda_2)}z_{23}^{i\lambda_1-1}z_{31}^{i\lambda_2-1} \label{3point}
\end{align}
Similarly, the four-point celestial gluon amplitude of the four-point MHV amplitude;
\begin{equation}
    A_{--++}=\delta^{(4)}(\omega_1q_1+\omega_2q_2-\omega_3q_3-\omega_4 q_4)\frac{\omega_1\omega_2}{\omega_3\omega_4}\frac{z_{12}^{3}}{z_{23}z_{34}z_{41}},
\end{equation}
takes the form \cite{StringOnCS}
\begin{align}
   \mathcal{A}_{--++} = &8\pi\delta(r-\bar{r}) z_{34}^{i\lambda_2-2}z_{12}^{-i\lambda_1} z_{24}^{i(\lambda_1+\lambda_3)}z_{23}^{-i(\lambda_2+\lambda_3)} \nonumber \\
    &\bar{z}_{34}^{i\lambda_1}\bar{z}_{12}^{-i\lambda_2-2}\bar{z}_{13}^{-i(\lambda_1+\lambda_3)}\bar{z}_{14}^{i(\lambda_2+\lambda_3)}\Theta(r-1)\delta(\sum_{j=1}^{4}\lambda_j) \label{4point}
\end{align}
where the cross ratio \(r\) is defined as
\begin{align}
    r \equiv \frac{z_{12} z_{34}}{z_{23} z_{41}} .
\end{align}
The cross ratio \(r\) is restricted to the region \(r > 1\) and satisfies the reality condition \(r = \bar{r}\).

The celestial graviton amplitudes can be constructed in a similar manner by applying the Mellin transformation to graviton scattering amplitudes. However, unlike the gluon case, the Mellin integrals for graviton amplitudes generally contain additional powers of energies. As a consequence, the resulting energy integrals are no longer convergent in the standard principal continuous series with $\Delta_j=1+i\lambda_j$.

For The three-point celestial graviton amplitudes take the form \cite{StringOnCS}
\begin{align}
    \mathcal{A}_{--,--,++} = 4 z_{21}^{2-i(\lambda_1+\lambda_2)}z_{23}^{i\lambda_1-1}z_{31}^{i\lambda_2-1}\delta(\bar{z}_{13})\delta(\bar{z}_{23})\int_{0}^{\infty}d\omega_3 \, \omega_{3}^{i\sum_{j=1}^3\lambda_j}
\end{align}
In contrast to the gluon amplitude in (\ref{3point}), the remaining energy integral is divergent. To regulate this divergence, one analytically continues the conformal dimension by shifting $\lambda_3$ to  $\lambda_3^{\prime}+i$ with $\lambda_3^{\prime} \in \mathbb{R}$ \cite{Puhm:2019zbl}. The energy integral can then be interpreted as a distribution, yielding
\begin{align}
    \mathcal{A}_{--,--,++} = 8\pi  z_{21}^{2-i(\lambda_1+\lambda_2)}z_{23}^{i\lambda_1-1}z_{31}^{i\lambda_2-1}\delta(\bar{z}_{13})\delta(\bar{z}_{23})\delta (\lambda_1+\lambda_2+\lambda_3^{\prime})
\end{align}
In the above expression, the Dirac delta function does not impose the constraint $\sum_{j=1}^{3}\lambda_j=0$, as in gluon case. Instead, the sum of $\lambda_j$ is required to be eqaul to the unit imagianry number, $i$.

Similarly, the four-point celestial graviton amplitudes take the form
\begin{align}
    \mathcal{A}_{--,--,++,++} = &8\pi(-1)^{i(\lambda_2+\lambda_3)}\delta(r- \bar{r})\Theta(r-1)z_{12}^{1-i(\lambda_1+\lambda_2)}z_{13}^{-1} \nonumber\\
    &z_{14}^{1+i\lambda_2}z_{23}^{-i\lambda_3}z_{24}^{i(\lambda_1+\lambda_3)}z_{34}^{-2}\bar{z}_{12} \bar{z}_{13}^{-1-i(\lambda_1+\lambda_3)}\bar{z}_{14}^{-3+i\lambda_3}\nonumber \\
    &\bar{z}_{23}^{-2-i\lambda_2}\bar{z}_{34}^{6+i(\lambda_1+\lambda_2)}\delta(\lambda_1+\lambda_2+\lambda_3+\lambda_4^{\prime}).
\end{align}
where $\lambda_4 = \lambda_4^{\prime}  +2i$. The above expression is obtained using a similar argument as in the three-point case to regulate the energy integral. In this case, the sum of $\lambda_j$ for the four-point amplitude is required to be $2i$. More generally, for the $n$-point graviton amplitudes, the same argument can be applied by analytically continuing $\lambda_n$ to $\lambda_n^{\prime}+(n-2)i$. Consequently, the constraint on the sum of $\lambda_j$ becomes $\sum_{j=1}^{n}\lambda_j = (n-2)i$.
%%%%%%%%%%%%%%%%%%%%%%%%%%%%%%%%%%%%%%%%%%%%%%%%%%%%%%%%%%%%%%%%%%%%%%%%%%%%%%%%%%%%%%%%%%%%%%%%%%%%%%%%%%%%%%%%%%%%%%%%%%%%%%%%%%%%%%%%%%%%%%%%%%%%%%%%%%%%%%%%%%%%%%%%
\section{Construction of conformally-invariant celestial amplitudes}\label{construct}
In this section, we construct a conformally invariant representation of celestial amplitudes. Although string amplitudes are defined on a different two-dimensional space, namely the worldsheet, they provide an important source of inspiration due to their exact conformal invariance. In particular, closed-string amplitudes can be expressed as worldsheet integrals of the form \cite{Green_Schwarz_Witten_2012}
\begin{equation}
    A^{\text{cl}}_n=\frac{1}{\text{Vol}}\int_{\mathbb{C}}\prod_{i=1}^n d^2z_i \prod_{1\leq i< j \leq n} |z_{ij}|^{\alpha'k_i\cdot k_j}z_{ij}^{n_{ij}}\bar{z}_{ij}^{\bar{n}_{ij}},
\end{equation}
 is conformally invariant. $n_{ij}, \bar{n}_{ij}\in \mathbb{Z}$ depend on the external string states. We emphasize that the complex coordinates $z_i,\bar{z}_i$ appearing here represent worldsheet insertion points and are unrelated to the celestial coordinates introduced earlier.

Motivated by this structure, our goal is to seek an appropriate transformation that maps the celestial variables $(z_j,\bar{z}_j)$ to the new set of variables $(s_j,\bar{s}_j)$ such that the new celestial amplitudes become conformally invariant.

\subsection{The integral transformation}
By inspecting the form of the string amplitudes, we propose the following integral transformation
\begin{equation}
    \phi:f_n(z_{ij},\bar{z}_{ij}) \longmapsto F_n(s_j,\bar{s}_j),
\end{equation}
such that 
\begin{align}
    F_n(s_j,\bar{s}_j) \equiv \int \prod_{j=1}^{n} d^2 z_j \prod_{0 \leq l < k \leq n} z_{lk}^{-1 + s_l + s_k}\bar{z}_{lk}^{-1 + \bar{s}_l + \bar{s}_k} \, f_n(z_{lk},\bar{z}_{lk}), \label{trans1}
\end{align}
where $z_{ij}= z_i-z_j$. Remember that the function $f_n(z_{lk},\bar{z}_{lk})$ depends only in the differences $z_{ij}$ and $\bar{z}_{ij}$.

For this map to be well defined, we require the existence of a corresponding inverse transformation. We postulate the inverse map to take the form
\begin{align}
    f_n(z_{ij},\bar{z}_{ij}) \equiv \mathcal{C}_n \int \prod_{j=1}^{n} ds_j \, d\bar{s}_j \prod_{0 \leq l < k \leq n} z_{lk}^{-1 - s_l - s_k} \, \bar{z}_{lk}^{-1 - \bar{s}_l - \bar{s}_k}F_n(s_j,\bar{s}_j), \label{inverse}
\end{align}
where $\mathcal{C}_n$ is a normalization factor to be determined. Consistency of the construction requires that the inverse transformation reproduces the original function $f_n$ when applied to $F_n$. To this extent, we start by exploring the lower-point mapping then generalizing it to the $n$-point case.

\subsection{Three-point case}
Applying the inverse map $\phi^{-1}$ to $F_3$ reads
\begin{align}
    f_3(z_{ij},\bar{z}_{ij}) = \mathcal{C}_3 \int \prod_{i=1}^{3} ds_i \, d\bar{s}_i \, 
    z_{12}^{-1 - s_1 - s_2} z_{13}^{-1 - s_1 - s_3} z_{23}^{-1 - s_2 - s_3} \, \nonumber \\
    \bar{z}_{12}^{-1 - \bar{s}_1 - \bar{s}_2} \bar{z}_{13}^{-1 - \bar{s}_1 - \bar{s}_3} \bar{z}_{23}^{-1 - \bar{s}_2 - \bar{s}_3}F_3(s_j,\bar{s}_j).
\end{align}
Substituting the explicit expression for $F_{3}(s_j,\bar{s}_j)$ from (\ref{trans1}) into the above equation yields
\begin{align}
    f_3(z_{ij},\bar{z}_{ij}) = \, &\int \prod_{i=1}^{3} d^2 z_i'\,\,\,\bigg{[} \mathcal{C}_3\int \prod_{i=1}^{3} ds_i \, d\bar{s}_i \, 
    z_{12}^{-1 - s_1 - s_2} z_{13}^{-1 - s_1 - s_3}  \nonumber \\
    &\times z_{23}^{-1 - s_2 - s_3}\bar{z}_{12}^{-1 - \bar{s}_1 - \bar{s}_2} \bar{z}_{13}^{-1 - \bar{s}_1 - \bar{s}_3} \bar{z}_{23}^{-1 - \bar{s}_2 - \bar{s}_3} z_{12}^{\prime -1 + s_1 + s_2}\nonumber \\
    &\times \, 
    z_{13}^{\prime -1 + s_1 + s_3} z_{23}^{\prime -1 + s_2 + s_3} \bar{z}_{12}^{\prime-1 + \bar{s}_1 + \bar{s}_2} \bar{z}_{13}^{\prime-1 +\bar{s}_1 + \bar{s}_3}
    \, \nonumber \\
    &\times  \bar{z}_{23}^{\prime-1 + \bar{s}_2 + \bar{s}_3}\bigg{]}f_3(z_{ij}',\bar{z}_{ij}'). \label{inv3}
\end{align}
The consistency of the inverse map requires that the integral inside the square brackets in \eqref{inv3} reduces to a product of delta functions, namely,
\begin{align}
    \bigg{[}\dots\bigg{]} \to \prod_{j=1}^{3}\delta(z'_j-z_j)\delta(\bar{z}'_j-\bar{z}_j). \label{eq}
\end{align}

\begin{figure}[ht]
    \centering
    \begin{tikzpicture}[scale=0.5]
        % Axes
        \draw[<->,thick] (-4,0) -- (4,0) node[right] {\large $\Re s_1$};
        \draw[<->,thick] (0,-4) -- (0,4) node[above] {\large $\Im s_1$};

        % Contour
        \draw[thick,blue] (2,-4) -- (2,4);

        % Arrows along contour
        \draw[->,thick,blue] (2,-4) -- (2,-1);
        \draw[->,thick,blue] (2,-1) -- (2,3);

        % Labels
        \node at (2.5,0.3) {\large $c_1$};
        \node at (4.5,-3) {\large $C_1 = c_1 + i y_1$};
    \end{tikzpicture}
    \caption{Contour of integration for $s_1$ in the complex plane.}
    \label{s1}
\end{figure}
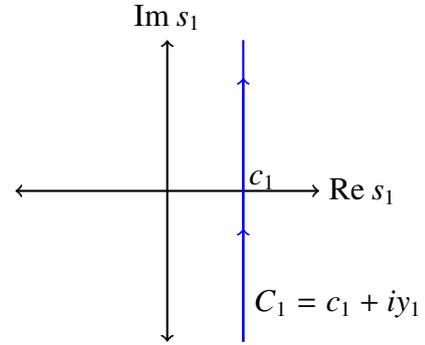

To demonstrate this explicitly, we begin by considering the integration over $s_1$, which appears in the square brackets of \eqref{eq}. The relevant integral is
\begin{align}
    I_{s_1} \equiv \int_{C_1} ds_1 \, 
    z_{12}^{\prime -\frac{1}{2}+s_1} z_{12}^{-\frac{1}{2}-s_1} 
    z_{13}^{\prime -\frac{1}{2}+s_1} z_{13}^{-\frac{1}{2}-s_1}. \label{s1-int}
\end{align}
To perform this integration, we use the contour of $s_1$ shown in Figure~\ref{s1} and write 
\begin{align}
    z_{ij}^d = \exp\big(d \ln z_{ij}\big).
\end{align}

After performing the contour integration over $s_1$, (\ref{s1-int}) becomes
\begin{align}
   I_{s_1} = 2\pi i \, z_{12}^{\prime -\frac{1}{2}+c_1} z_{12}^{-\frac{1}{2}-c_1} 
   z_{13}^{\prime -\frac{1}{2}+c_1} z_{13}^{-\frac{1}{2}-c_1} 
   \, \delta\Bigg(\ln \frac{z_{12} z_{13}}{z_{12}^{\prime} z_{13}^{\prime}}\Bigg).
\end{align}
Applying the standard property of the delta function,
\begin{align}
     \delta(f(x)) = \sum_i \frac{\delta(x-x_i)}{|f^{\prime}(x_i)|}, \label{delta}
\end{align}
we then find
\begin{align}
    I_{s_1} = 2\pi i \, z_{12}^{\prime -\frac{1}{2}+c_1} z_{12}^{-\frac{1}{2}-c_1} 
    z_{13}^{\prime -\frac{1}{2}+c_1} z_{13}^{-\frac{1}{2}-c_1} 
    \, z_{12} z_{13} \, \delta\big(z_{12}^{\prime} z_{13}^{\prime} - z_{12} z_{13}\big).
\end{align}
Similar results can be obtained for the remaining integration with respect to variables $s_j$ and $\bar{s}_j$ with $j=2,3$. Altogether, these integrations yield six delta functions,
\begin{align}
  &\delta(z_{12}^{\prime} z_{13}^{\prime} - z_{12} z_{13}) \, 
          \delta(z_{12}^{\prime} z_{23}^{\prime} - z_{12} z_{23}) \, 
          \delta(z_{13}^{\prime} z_{23}^{\prime} - z_{13} z_{23}) \nonumber \\
      &\times \delta(\bar{z}_{12}^{\prime} \bar{z}_{13}^{\prime} - \bar{z}_{12} \bar{z}_{13}) \, 
          \delta(\bar{z}_{12}^{\prime} \bar{z}_{23}^{\prime} - \bar{z}_{12} \bar{z}_{23}) \, 
          \delta(\bar{z}_{13}^{\prime} \bar{z}_{23}^{\prime} - \bar{z}_{13} \bar{z}_{23}).\label{dirac deltas}
\end{align}
Ultimately, our goal is to express these constraints in the form $\prod_{j}\delta(z'_j - z_j) \, \delta(\bar{z}'_j - \bar{z}_j)$. Let us first focus on the holomorphic sector, i.e. the delta functions appearing
in the first line of \eqref{dirac deltas}. At first glance, the three constraints imply the equalities $z_{ij}=z'_{ij}$ for all pairs $(i,j)$. This is problematic when one attempts to express the variables $z_{12}, z_{13}, z_{23}$ in terms of $z_1,z_2,z_3$, which causes a divergence as the variables $z_{ij}$ are not
all independent leading to overcounting the degrees of freedom.

To deal with this, we rewrite the three constraints in
terms of a set of linearly independent variables in which we choose to be $z_{12}$ and $z_{13}$. Accordingly, we may obtain
\begin{align}
     \delta(z_{12}^{\prime} z_{13}^{\prime} - z_{12} z_{13}) \, 
          \delta(z_{12}^{\prime} z_{23}^{\prime} - z_{12} z_{23}) \, 
          \delta(z_{13}^{\prime} z_{23}^{\prime} - z_{13} z_{23}) \nonumber \\
          =  \frac{1}{\det J_3 \big\vert_{z^{\prime}_{ij}=z_{ij}}} 
    \, \delta(z_{12}^{\prime} - z_{12}) \, \delta(z_{13}^{\prime} - z_{13}),
\end{align}
where $J_3$ denotes the Jacobian matrix associated with this change of variables,
\begin{align}
    J_3 = \begin{pmatrix}
        z_{13} & z_{13} \\
        z_{13}-2z_{12} & z_{12} \\
        -z_{13} & 2z_{13}-z_{12}
    \end{pmatrix}.
\end{align}

Since $J_3$ is not a square matrix, the standard determinant is not defined. Instead, we employ the generalized determinant for a non-square matrix $J$ of size $m \times n$ with $m>n$, which is defined as \cite{Hannah01051996}
\begin{align}
    \det(J) \equiv \sqrt{J^{T} J}.
\end{align}
The determinant of $J_3$ is 
\begin{align}
    \det(J_3) = \sqrt{z_{12}^2-z_{13} z_{12}+z_{13}^2}.
\end{align}
To proceed, we insert the identity
\begin{align}
    1 = \int dy \, \delta(z'_1 - z_1 - y). \label{iden}
\end{align}
This allows us to rewrite the delta function as
\begin{align}
      &\delta(z_{12}^{\prime} z_{13}^{\prime} - z_{12} z_{13}) \, 
          \delta(z_{12}^{\prime} z_{23}^{\prime} - z_{12} z_{23}) \, 
          \delta(z_{13}^{\prime} z_{23}^{\prime} - z_{13} z_{23}) \nonumber \\
          &= \frac{1}{\det J_3 \big\vert_{z^{\prime}_{ij} = z_{ij}}} 
    \int dy \, \delta(z'_1 - z_1 - y) \, \delta(z'_2 - z_2 - y) \, \delta(z'_3 - z_3 - y).
\end{align}
A similar expression holds for the anti-holomorphic sector,
\begin{align}
    &\delta(\bar{z}_{12}^{\prime} \bar{z}_{13}^{\prime} - \bar{z}_{12} \bar{z}_{13}) \, 
          \delta(\bar{z}_{12}^{\prime} \bar{z}_{23}^{\prime} - \bar{z}_{12} \bar{z}_{23}) \, 
          \delta(\bar{z}_{13}^{\prime} \bar{z}_{23}^{\prime} - \bar{z}_{13} \bar{z}_{23})\nonumber \\
          &= \frac{1}{\det \bar{J}_3 \big\vert_{\bar{z}^{\prime}_{ij} = \bar{z}_{ij}}} 
    \int dw \, \delta(\bar{z}'_1 - \bar{z}_1 - w) \, \delta(\bar{z}'_2 - \bar{z}_2 - w) \, \delta(\bar{z}'_3 - \bar{z}_3 - w).
\end{align}
Since the function $f_3(z_{ij}, \bar{z}_{ij})$ depends only on the differences $z_{ij}$ and $\bar{z}_{ij}$, it is invariant under global shifts. As a result, the integrations over the global parameters $y$ and $w$ in
\eqref{inv3} factorize, yielding
\begin{align}
    \left(\int dy \, dw\right)
    \int \prod_{j=1}^{3} d^2 z'_j \, 
    \prod_{l=1}^{3} \delta(z'_l - z_l) \, \delta(\bar{z}'_l - \bar{z}_l) 
    \, f_3(z'_{ij}, \bar{z}'_{ij}), \label{finaldelta}
\end{align}
The integrations over \(y\) and \(w\) yield divergences as expected. However, these divergences can be
consistently absorbed into the normalization constant
\(\mathcal{C}_3\). Substituting (\ref{finaldelta}) and all factors into (\ref{inv3}), we obtain
\begin{align}
    f_3(z_{ij},\bar{z}_{ij}) = 
    &\frac{ \mathcal{C}_3 (2\pi i)^6 \int dy \, dw }{\det J_3\big|_{z'_{ij} = z_{ij}}\det \bar{J}_3\big|_{\bar{z}'_{ij} = \bar{z}_{ij}}}  
    \int \prod_{i=1}^3 d^2z'_i \, 
     \prod_{l=1}^3 \delta(z'_l-z_l)\nonumber \\
     &\times\delta(\bar{z}'_l-\bar{z}_{l}) f_3(z'_{ij},\bar{z}'_{ij}) 
     z_{12}^{\prime-\frac{1}{2}+c_1} z_{12}^{-\frac{1}{2}-c_1}z_{13}^{\prime-\frac{1}{2}+c_1}z_{13}^{-\frac{1}{2}-c_1} \nonumber \\
     &\times z_{12}z_{13}\, z_{12}^{\prime-\frac{1}{2}+c_2} z_{12}^{-\frac{1}{2}-c_2}z_{23}^{\prime-\frac{1}{2}+c_2}z_{23}^{-\frac{1}{2}-c_2} \nonumber  \\ 
    &\times z_{12}z_{23} z_{23}^{\prime-\frac{1}{2}+c_3} z_{23}^{-\frac{1}{2}-c_3}z_{13}^{\prime-\frac{1}{2}+c_3}z_{13}^{-\frac{1}{2}-c_3} \nonumber \\
    &\times z_{23}z_{13}\,\bar{z}_{12}^{\prime-\frac{1}{2}+\bar{c}_1} \bar{z}_{12}^{-\frac{1}{2}-\bar{c}_1}\bar{z}_{13}^{\prime-\frac{1}{2}+\bar{c}_1}\bar{z}_{13}^{-\frac{1}{2}-\bar{c}_1} \nonumber  \\ 
     &\times\bar{z}_{12}\bar{z}_{13} \bar{z}_{12}^{\prime-\frac{1}{2}+\bar{c}_2} \bar{z}_{12}^{-\frac{1}{2}-\bar{c}_2}\bar{z}_{23}^{\prime-\frac{1}{2}+\bar{c}_2}\bar{z}_{23}^{-\frac{1}{2}-\bar{c}_2} \nonumber \\
     &\times\bar{z}_{12}\bar{z}_{23} \,\bar{z}_{13}^{\prime-\frac{1}{2}+\bar{c}_3} \bar{z}_{13}^{-\frac{1}{2}-\bar{c}_3}\bar{z}_{23}^{\prime-\frac{1}{2}+\bar{c}_3}\bar{z}_{23}^{-\frac{1}{2}-\bar{c}_3} \bar{z}_{13}\bar{z}_{23}.
\end{align}
Performing the integrations over $z'$ and $\bar{z}'$. This gives
\begin{align}
     f_3(z_{ij},\bar{z}_{ij}) &= 
     \frac{ \mathcal{C}_3 (2\pi i)^6 \mathrm{Vol} }{\det J_3\big|_{z'_{ij} = z_{ij}}\det \bar{J}_3\big|_{\bar{z}'_{ij} = \bar{z}_{ij}}}  
     f_3(z_{ij},\bar{z}_{ij}).
\end{align}
where \(\mathrm{Vol}\) denotes the divergent volume factor arising from the integrations over the translational parameters $y$ and $w$. For the inverse transformation to be well-defined, the overall coefficient must be unity. 
Hence, we find
\begin{align}
   \mathcal{C}_3 = \frac{\det J_3\big|_{z'_{ij} = z_{ij}}\det \bar{J}_3\big|_{\bar{z}'_{ij} = \bar{z}_{ij}}}{(2\pi i)^6 \, \mathrm{Vol}}.
\end{align}
With this choice of normalization, the inverse transformation takes the explicit form
\begin{align}
      f_3(z_{ij},\bar{z}_{ij}) 
      &= \frac{\det J_3\det \bar{J}_3}{(2\pi i)^6 \, \mathrm{Vol}}\bigg|_{\substack{z'_{ij} = z_{ij}\\\bar{z}'_{ij} = \bar{z}_{ij}}}
      \int \prod_{i=1}^{3} ds_i \, d\bar{s}_i \,
      z_{12}^{-1 - s_1 - s_2} 
      z_{13}^{-1 - s_1 - s_3} 
      \nonumber \\
      &\times z_{23}^{-1 - s_2 - s_3} \, 
        \bar{z}_{12}^{-1 - \bar{s}_1 -\bar{s}_2} 
      \bar{z}_{13}^{-1 - \bar{s}_1 - \bar{s}_3} 
      \bar{z}_{23}^{-1 - \bar{s}_2 - \bar{s}_3}
      F_3(s_j,\bar{s}_j).
\end{align}

\subsubsection{Four-point case}
In the four-point case, there are a total of eight delta functions. For the holomorphic part, we obtain four delta functions imposing four constraints. However, only three independent variables exist, which we may choose to be $z_{12}$, $z_{13}$, and $z_{14}$. As in the three-point case, this may cause divergences that must be treated carefully. Proceeding analogously, the holomorphic part of the four-point delta functions can be written as
\begin{align}
     \frac{1}{\det J_4}\bigg{\vert}_{z'_{ij}=z_{ij}} \delta(z'_{12}-z_{12})\delta(z'_{13}-z_{13})\delta(z'_{14}-z_{14})
\end{align}
Similarly, for the anti-holomorphic sector, we obtain
\begin{align}
       \frac{1}{\det \bar{J}_4}\bigg{\vert}_{\bar{z}'_{ij}=\bar{z}_{ij}} \delta(\bar{z}'_{12}-\bar{z}_{12})\delta(\bar{z}'_{13}-\bar{z}_{13})\delta(\bar{z}'_{14}-\bar{z}_{14})
\end{align}
where $\det J_4$ and $\det \bar{J}_4$ denote the Jacobians associated with the change of variables. To expose the redundancy, we insert the identity (\ref{iden}). This yields
\begin{align}
     \frac{1}{\det J_4}\bigg{\vert}_{z'_{ij}=z_{ij}} \int dy\,\delta(z'_{1}-z_{1}-y)\delta(z'_{2}-z_{2}-y)\nonumber \\
    \delta(z'_{3}-z_{3}-y)\delta(z'_{4}-z_4-y)
\end{align}
and 
\begin{align}
     \frac{1}{\det J_4}\bigg{\vert}_{z'_{ij}=z_{ij}} \int dw\,\delta(\bar{z}'_{1}-\bar{z}_{1}-w)\delta(\bar{z}'_{2}-\bar{z}_{2}-w)\nonumber \\
    \delta(\bar{z}'_{3}-\bar{z}_{3}-w)\delta(\bar{z}'_{4}-\bar{z}_4-w)
\end{align} 
Since the function $f_4(z_{ij},\bar{z}_{ij})$ depends only on the $z_{ij}$ and $\bar{z}_{ij}$, the global shifts in $z_i$ and $\bar{z}_i$ parameterized by $y$ and $w$ do not affect their values. Consequently, the integrations over $y$ and $w$ factorize. This allows us to fix the normalization constant $\mathcal{C}_4$ to be
\begin{align}
    \mathcal{C}_4 = \frac{\det J_4\big|_{z'_{ij} = z_{ij}}\det \bar{J}_4\big|_{\bar{z}'_{ij} = \bar{z}_{ij}}}{(2\pi i)^{8} \, \mathrm{Vol}}.
\end{align}
With this normalization, the inverse transformation for the four-point case is given by
\begin{align}
    f_4(z_{ij},\bar{z}_{ij}) = \frac{\det J_4\det \bar{J}_4}{(2\pi i)^{8} \, \mathrm{Vol}}\bigg|_{\substack{z'_{ij} = z_{ij}\\\bar{z}'_{ij} = \bar{z}_{ij}}} \int \prod_{j=1}^{4} ds_i d\bar{s}_i \nonumber \\
    \times \prod_{0\leq l < k \leq 4} z_{lk}^{-1-s_l-s_k} \bar{z}_{lk}^{-1-\bar{s}_l-\bar{s}_k} F_{4}(s_n,\bar{s}_n).
\end{align}

By induction, this procedure can be extended straightforwardly to the general $n$-point case. The resulting inverse transformation takes the form given in \eqref{inverse}, with the normalization factor $\mathcal{C}_n$ given by
\begin{align}
    \mathcal{C}_n = \frac{\det J_n\big|_{z'_{ij} = z_{ij}}\det \bar{J}_n\big|_{\bar{z}'_{ij} = \bar{z}_{ij}}}{(2\pi i)^{2n} \, \mathrm{Vol}}.
\end{align}

\section{Conformal invariance of the new celestial amplitudes} 
\label{TestFor}
In this section, we examine whether the transformed celestial amplitudes are conformally invariant or not. Three- and Four-point case condition were discussed. Then we generalize the result to the $n$-point case.
\subsection{Celestial gluon amplitudes}
\subsubsection{Three-point case}
Applying the integral transformation (\ref{trans1}) to the MHV three-point celestial amplitudes (\ref{3point}), we obtain
\begin{align}
    \tilde{\mathcal{A}}_3 \equiv 8\pi\delta(\sum_{j=1}^{3}\lambda_j)(-1)^{-i\lambda_1}\int d^2z_1d^2z_2d^2z_3\, z_{12}^{s_1+s_2-i(\lambda_1+\lambda_2)}\delta(\bar{z}_{13})\nonumber\\ \times\delta(\bar{z}_{23})z_{13}^{-2+s_1+s_3+i\lambda_2} z_{23}^{-2+s_2+s_3+i\lambda_1} \bar{z}_{12}^{-1+\bar{s}_1+\bar{s}_2}\bar{z}_{13}^{-1+\bar{s}_1+\bar{s}_3}\bar{z}_{23}^{-1+\bar{s}_2+\bar{s}_3}
\end{align}
Then we perform the change of variables:
\begin{align}
    z_{j} = \prod_{l=1}^{j}w_l \quad \text{and} \qquad \bar{z}_{j} = \prod_{l=1}^{j}\bar{w}_l. \label{changew}
\end{align}
The Jacobian of the transformation is $w_1^2w_2\bar{w}_1^2\bar{w}_2$. In this variable, the integration over $(w_1,\bar{w}_1)$ factorizes. Let us focus on this part of the integral:
\begin{align}
    I_1 &\equiv \int d^2w_1 \,w_1^{-2+2\sum_{j=1}^{3}s_j} \bar{w}_1^{-3+2\sum_{j=1}^3\bar{s}_j} \nonumber \\
    &= \int d^2w_1\,|w_1|^{-4}w_1^{2\sum_{j=1}^3s_j} \bar{w}_1^{-1+2\sum_{j=1}^2\bar{s}_j}. \label{I1}
\end{align}
To compute this integral, we employ the Schwinger parametrization
\begin{align}
    |z|^{2a-2} = \frac{1}{\Gamma(1-a)} \int_{0}^{\infty} dt \,\,t^{-a} e^{-|z|^2t}. \label{schwinger}
\end{align}
This yields
\begin{align}
    I_1 = \int_{0}^{\infty}dt\,\,t \int d^2w_1\,\, e^{-|w_1|^2t} w_1^{2\sum_{j=1}^3s_j} \bar{w}_1^{-1+2\sum_{j=1}^2\bar{s}_j}
\end{align}
Introducing source terms $(J,\bar{J})$, the integral can be written as 
\begin{align}
    I_1 = \int_{0}^{\infty}dt\,\,t \left(\frac{\delta}{\delta J}\right)^{\bar{m}}\left(\frac{\delta}{\delta \bar{J}}\right)^{m} \int d^2w_1 e^{-|w_1|^2t+\bar{w}_1J+w_1\bar{J}} \bigg{\vert}_{J=\bar{J} =0} \label{funcderiv}
\end{align}
where \[m=2\sum_{j=1}^3s_j \quad \text{and} \quad \bar{m}=-1+ 2\sum_{j=1}^3\bar{s}_j.\] 
Using the Gaussian integral and performing the functional derivatives with respect to \( J \) and \( \bar{J} \), together with the fact that the result is non-vanishing only when \( \bar{m} = m \), we obtain
\begin{align}
    I_1 = 2\pi i m!\int_{0}^{\infty}dt\,\,t^{-m} 
\end{align}
 This result is obvious when $m$ and $\bar{m}$ are integers. However, it can be extended to the non-integer case using the fractional derivatives discussed in \ref{Fractional}. To perform the $t$-integration, we change the variables
\begin{align}
    t = e^{x}
\end{align}
which yields
\begin{align}
    I_1 = (2\pi i)^2 m!\delta(m-1).
\end{align}
Consequently, we obtain the constraints
\begin{align}
    \sum_{j=1}^{3} s_j = \frac{1}{2} \hspace{0.5cm} \text{and} \hspace{0.5cm}\sum_{j=1}^3 \bar{s}_j = 1. \label{cond3point}
\end{align}
We now demonstrate that these conditions are necessary to ensure conformal invariance of the transformed amplitude. Consider the scaling transformation
\begin{align}
    z_j \to \alpha z_j, \qquad \bar{z}_j \to \bar{\alpha}\bar{z}_j \label{scaling}
\end{align}
Under this transformation, the amplitudes $\tilde{\mathcal{A}}_3$ transform as 
\begin{align}
    \tilde{\mathcal{A}}^{\prime}_3 = \alpha^{-1+2\sum_{j=1}^3s_j -\frac{i}{2}\sum_{j=1}^3\lambda_j} \bar{\alpha}^{-2+2\sum_{j=1}^3\bar{s}_j -\frac{i}{2}\sum_{j=1}^3\lambda_j} \tilde{\mathcal{A}}_3
\end{align}
Applying the condition in (\ref{cond3point}) together with the constraint $\sum_{j=1}^{3} \lambda_j = 0$, we find 
\begin{align}
    \tilde{\mathcal{A}}^{\prime}_3 = \tilde{\mathcal{A}}_3
\end{align}
This shows that the transformed amplitude \(\tilde{\mathcal{A}}_3^{\prime}\) is invariant under the conformal transformations.

\subsubsection{Four-point case}
For the four-point case, we apply the integral transformation to (\ref{4point}) and obtain
\begin{align}
  \tilde{\mathcal{A}}_4 = &8\pi\delta(\sum_{j=1}^{4}\lambda_j) \int d^2z_1d^2z_2d^2z_3d^2z_4 \,\, z_{12}^{-1+s_1+s_2-i\lambda_1} z_{13}^{-1+s_1+s_3} \nonumber \\ &z_{14}^{-1+s_1+s_4}z_{23}^{-1+s_2+s_3-i(\lambda_2+\lambda_3)}z_{24}^{-1+s_2+s_4+i(\lambda_1+\lambda_3)}  z_{34}^{-3+s_3+s_4+i\lambda_2} \nonumber \\
    &\bar{z}_{12}^{-3+\bar{s}_1+\bar{s}_2-i\lambda_2}\bar{z}_{13}^{-1+\bar{s}_1+\bar{s}_3-i(\lambda_1+\lambda_3)}\bar{z}_{14}^{-1+\bar{s}_1+\bar{s}_4+i(\lambda_2+\lambda_3)}\bar{z}_{23}^{-1+\bar{s}_2+\bar{s}_3}\nonumber \\
    &\bar{z}_{24}^{-1+\bar{s}_2+\bar{s}_4}\bar{z}_{34}^{-1+\bar{s}_3+\bar{s}_4+i\lambda_1}\Theta(r-1)\delta(r-\bar{r})
\end{align}
Using the change of variables \eqref{changew}, the integration over $w_1$ factorizes and reduces to
\begin{align}
   I_2 \equiv \int d^2 w_1 |w_1|^{-4} w_1^{-3+3\sum_{j=1}^4s_j}\bar{w}_1^{-3+3\sum_{j=1}^{4}\bar{s}_j}. \label{I2}
\end{align}
This integral evaluates to 
\begin{align}
    I_2 = (2\pi i)^2 m!\delta(m-1)
\end{align}
where 
\begin{equation}
m= -3+3\sum_{j=1}^4s_j = \bar{m} = -3+3\sum_{j=1}^{4}\bar{s}_j.
\end{equation} Consequently, the emerging conditions for the four-point case are
\begin{align}
     \sum_{j=1}^4 s_j =\frac{4}{3} \hspace{0.5cm}\text{and}\hspace{0.5cm} \sum_{j=1}^{4}\bar{s}_j = \frac{4}{3}. \label{cond4point}
\end{align}
Under the scaling transformation in (\ref{scaling}), the transformed amplitude $\tilde{\mathcal{A}}_4$ transforms as 
\begin{align}
    \tilde{\mathcal{A}}_4^{\prime} = \alpha^{-4+3\sum_{j=1}^4s_j -\frac{i}{2}\sum_{j=1}^4\lambda_j} \bar{\alpha}^{-4+3\sum_{j=1}^4\bar{s}_j -\frac{i}{2}\sum_{j=1}^4\lambda_j} \tilde{\mathcal{A}}_4.
\end{align}
Imposing the condition \eqref{cond4point} together with the constraint $\sum_{j=1}^{4} \lambda_j = 0$, we find
\begin{align}
     \tilde{\mathcal{A}}_4^{\prime} = \tilde{\mathcal{A}}_4
\end{align}
Therefore, the transformed four-point celestial amplitude is invariant under conformal transformations.
\subsubsection{$n$-point case}
For the \(n\)-point case, the transformed amplitude is given by
\begin{align}
    \tilde{\mathcal{A}}_n
    = \int \prod_{j=1}^{n} d^2 z_j
    \prod_{0 \leq l < k \leq n}
    z_{lk}^{-1 + s_l + s_k}
    \bar{z}_{lk}^{-1 + \bar{s}_l + \bar{s}_k}
    \, \mathcal{A}_{--+\dots+} 
\end{align}
where the \(n\)-point celestial MHV gluon amplitude \(\mathcal{A}_{--+\dots+}\) takes the form 
\begin{align}
    \mathcal{A}_{--+\dots+}
    =&\int_{0}^\infty \prod_{j=1}^n d\omega_j \omega_j^{i\lambda_j} \delta^{(4)}(\sum_{l=1}^n\epsilon_l\omega_l q_l) \nonumber \\
    &\times \frac{\omega_1\omega_2}{\omega_3\omega_4\ldots\omega_n}\frac{z_{12}^{3}}{z_{23} z_{34} z_{45} \dots z_{n1}} \label{n-point celestial amp} 
\end{align}
Using the decomposition of the Dirac delta functions in \cite{npointDelta}, we can rewrite (\ref{n-point celestial amp}) as 
\begin{align}
&\frac{i}{4}
    \frac{(1 - t_4)(1 - \bar{t}_4)}{t_4 - \bar{t}_4}
    \frac{z_{12}^{3}}{z_{23} z_{34} z_{45} \dots z_{n1}
    |z_{14}|^2 |z_{23}|^2}
    \nonumber \\
    & \times\int_{0}^{\infty}
    \prod_{j=5}^{n} d\omega_j \,
    \omega_j^{i\lambda_j - 1} (\omega_1^{*})^{i\lambda_1 + 1}
    (\omega_2^{*})^{i\lambda_2 + 1}
    (\omega_3^{*})^{i\lambda_3 - 1}
    (\omega_4^{*})^{i\lambda_4 - 1}\label{npoint}
\end{align}
where the variables \(t_4\), \(\bar{t}_4\), and \(\omega_j^{*}\) are given in \ref{nDelta}. To proceed, we apply the change of variables (\ref{changew}) from $(z_i,\bar{z}_i)$ to $(w_i,\bar{w}_i)$. Note that in this frame, $t_4$ and $\omega_i^{*}$ are $w_1$-independent, hence,  integration over $w_1$ can be factorized. The integration with respect to $w_1$ takes the form
\begin{align}
    I_3 \equiv &\int d^2 w_1 \, |w_1|^{-4}
    \, w_1^{3-\frac{n(n-1)}{2}+(n-1)\sum_{j=1}^{n}s_j}
    \nonumber \\
    &\times
    \bar{w}_1^{-1-\frac{n(n-3)}{2}+(n-1)\sum_{j=1}^{n}\bar{s}_j}.
\end{align}
Evaluating the above integral, we obtain
\begin{align}
    I_3 = (2\pi i)^2 m! \, \delta(m-1),
\end{align}
where
\begin{equation}
m = 3-\frac{n(n-1)}{2}+(n-1)\sum_{j=1}^{n}s_j
= \bar{m} = -1-\frac{n(n-3)}{2}+(n-1)\sum_{j=1}^{n}\bar{s}_j .    
\end{equation}

Accordingly, the conditions for the 
$n$-point case are given by
\begin{align}
     \sum_{j=1}^{n}s_j = \frac{n^2-n-4}{2(n-1)}
     \hspace{0.5cm} \text{and} \hspace{0.5cm}
     \sum_{j=1}^{n}\bar{s}_j = \frac{n^2-3n+4}{2(n-1)} .
     \label{condNpoint}
\end{align}

Under these conditions, the transformed $n$-point amplitude
$\tilde{\mathcal{A}_n}$ is likewise invariant under global conformal transformations.
\subsection{Celestial graviton amplitudes}
\subsubsection{Three-point case}
For the three-point graviton case, the transformed amplitudes take the form
\begin{align}
    \tilde{\mathcal{A}}_{3}^{G} = &8\pi\int d^2z_1\, d^2z_2 \, d^2z_3 \delta (\bar{z}_{13})\delta(\bar{z}_{23})\delta(\lambda_1+\lambda_2+\lambda_3^{\prime}) \nonumber \\
    &z_{12}^{1+s_1+s_2-i(\lambda_1+\lambda_2)}z_{13}^{-2+s_1+s_3+i\lambda_2} z_{23}^{-2+s_2+s_3+i\lambda_1}\nonumber \\
&\bar{z}_{12}^{-1+\bar{s}_1+\bar{s}_2}\bar{z}_{13}^{-1\bar{s}_1+\bar{s}_3}\bar{z}_{23}^{-1\bar{s}_2+\bar{s}_3}
\end{align}
Using the same procedure as in the gluon case, the integration over $w_1$ can be factorized and takes the form
\begin{align}
    I_4 \equiv \int d^2w_1 |w_1|^{-4} w_1^{1+2\sum_{j=1}^3s_j}\bar{w}_1
^{-1+2\sum_{j=1}^{n}\bar{s}_j}\end{align}
The above integral yields
\begin{align}
    I_4 = (2\pi i)^2 m! \, \delta(m-1),
\end{align}
where
\begin{align}
    m = 1+2\sum_{j=1}^3s_j = \bar{m} = -1+2\sum_{j=1}^{n}\bar{s}_j
\end{align}
Consequently, the conditions for the three-point graviton case are given by
\begin{align}
    \sum_{j=1}^3s_j = 0 \qquad \text{and} \qquad \sum_{j=1}^{3}\bar{s}_j = 1 \label{3condg}
\end{align}
Under the scaling transformation (\ref{scaling}), the transformed $\tilde{\mathcal{A}}^{G}_{3}$ transform as 
\begin{align}
    \tilde{\mathcal{A}}^{\prime \,G}_{3} = \alpha^{-\frac{1}{2}+2\sum_{j=1}^{3}s_j-\frac{i\sum_{j=1}^{3} \lambda_j}{2}}\bar{\alpha}^{-\frac{5}{2}+2\sum_{j=1}^3 \bar{s}_j-\frac{i\sum_{j=1}^{3} \lambda_j}{2}}\tilde{\mathcal{A}}^{G}_{3}
\end{align}
Applying the conditions in (\ref{3condg}) together with $\sum_{j=1}^3\lambda_j = i$, we obtain
\begin{align}
     \tilde{\mathcal{A}}^{\prime \,G}_{3} = \tilde{\mathcal{A}}^{G}_{3}
\end{align}
This shows that, under the integral transformation, the three-point graviton amplitude is invariant under global conformal transformations.
\subsubsection{Four-point case}
The four-point transformed graviton amplitude takes the form
\begin{align}
    \tilde{\mathcal{A}}_4^{G} = &8\pi (-1)^{i(\lambda_2+\lambda_3)}\delta(\lambda_1+\lambda_2+\lambda_3+\lambda_4^{\prime})\int \, d^2z_1d^2z_2d^2z_3d^2z_4 \nonumber \\
    &\delta(r-\bar{r}) \Theta(r-1) z_{12}^{s_1+s_2-i(\lambda_1+\lambda_2)}z_{13}^{-2+s_1+s_3}z_{14}^{s_1+s_4+i\lambda_2}  z_{23}^{-1+s_2+s_3-i\lambda_3} \nonumber\\
&z_{24}^{-1+s_2+s_4+i(\lambda_1+\lambda_3)}z_{34}^{-3+s_3+s_4} \bar{z}_{12}^{\bar{s}_1+\bar{s}_2}\bar{z}_{13}^{-2+\bar{s}_1+\bar{s}_3-i(\lambda_1+\lambda_3)}\bar{z}_{14}^{-4\bar{s}_1+\bar{s}_4+i\lambda_3}\nonumber\\
&\bar{z}_{23}^{-3+\bar{s}_2+\bar{s}_3-i\lambda_2}\bar{z}_{34}^{5+\bar{s}_3+\bar{s}_4+i(\lambda_1+\lambda_2)}
\end{align}
Using the change of variables in (\ref{changew}), the integration over $w_1$ takes the form
\begin{align}
    I_5 \equiv \int d^2w_1 \, |w_1|^{-4}w_1^{-2+3\sum_{j=1}^{4}s_j}\bar{w}_1^{-2+3\sum_{j=1}^{4}\bar{s}_j}
\end{align}
Evaluating this integral, we obtain 
\begin{align}
    I_5 = (2\pi i)^2 m! \, \delta(m-1),
\end{align}
where 
\begin{align}
    m = -2+3\sum_{j=1}^{n}s_j = \bar{m} =-2+3\sum_{j=1}^{n}\bar{s}_j
\end{align}
Therefore, the conditions for the four-point graviton case are given by
\begin{align}
    \sum_{j=1}^4 s_j = 1 \qquad \text{and} \qquad \sum_{j=1}^4 \bar{s}_j = 1 \label{4condg}
\end{align}
Under these conditions, together with $\sum_{j=1}^4\lambda_j = 2i$, the transformed amplitude is invariant under global conformal transformations.
%%%%%%%%%%%%%%%%%%%%%%%%%%%%%%%%%%%%%%%%%%
\subsubsection{$n$-point case}
For the $n$-point graviton case, the transformed amplitude is given by 
\begin{align}
    \tilde{\mathcal{A}}_n^{G} = \int \prod_{j=1}^{n} d^2 z_j
    \prod_{1 \leq l < k \leq n}
    z_{lk}^{-1 + s_l + s_k}
    \bar{z}_{lk}^{-1 + \bar{s}_l + \bar{s}_k}
    \, \mathcal{A}_{--,--,++,\dots,++} 
\end{align}
By employing the Mellin transformation, the $n$-point celestial MHV graviton amplitude \(\mathcal{A}_{--,++,\dots,++}\) takes the form 
\begin{align}
    \mathcal{A}_{--,++,\dots,++}
    =&\int_{0}^\infty \prod_{j=1}^n d\omega_j \omega_j^{i\lambda_j} \delta^{(4)}(\sum_{l=1}^n\epsilon_l\omega_l q_l) \nonumber \\
    &\sum_{P(i_3,i_4,\dots,i_n)}s_{1i_n}\left(\prod_{k=1}^{n-1} \beta_k \right) \left(A^{\text{YM}}_{1^-,2^-,i^+_3,i^+_4,\dots,i^+_n}\right)^2
    \label{n-point graviton amp} 
\end{align}
Using the same method as in the lower-point cases, we employ the change of variables in (\ref{changew}) and decompose the Dirac delta function as in \ref{nDelta}. The integration over $w_1$ can then be factorized and takes the form
\begin{align}
    I_6 \equiv \int d^2w_1 \, |w_1|^{-4}w_1^{\frac{-n^2+n+8}{2}+(n-1)\sum_{j=1}^{n}s_j}\bar{w}_1^{\frac{-n^2+5n-8}{2}+(n-1)\sum_{j=1}^{n}\bar{s}_j}
\end{align}
Performing the integration over $w_1$, we obtain
\begin{align}
    I_6 = (2\pi i)^2 m! \, \delta(m-1)
\end{align}
where 
\begin{align}
    m&= \frac{-n^2+n+8}{2}+(n-1)\sum_{j=1}^{n}s_j \nonumber \\
    &=\bar{m} = \frac{-n^2+5n-8}{2}+(n-1)\sum_{j=1}^{n}\bar{s}_j
\end{align}
Therefore, the conditions for $n$-point graviton case are given by 
\begin{align}
    \sum_{j=1}^n s_j = \frac{n^2-n-6}{2(n-1)} \qquad \text{and} \qquad \sum_{j=1}^n\bar{s}_j = \frac{n^2-5n+10}{2(n-1)} \label{condNgrav}
\end{align}
Under these conditions, together with $\sum_{j=1}^{n} \lambda_j = (n-2)i$, the transformed $n$-point graviton amplitude is invariant under the global conformal transformations.
\section{The physical interpretation of new variables $(s_j,\bar{s}_j)$}\label{comment}
In this section, we explore the relations between new variables, $(s_j,\bar{s}_j)$, and the physical data of the external particles, $(\Delta_j,J_j)$. Let us consider the transfromed $n$-point celestial amplitudes
\begin{align}
    \tilde{\mathcal{A}}_n(s_j,\bar{s}_j) = \int \prod_{j=1}^{n}d^2z_j \prod_{0 \leq l < k \leq n} z_{lk}^{-1 + s_l + s_k}\bar{z}_{lk}^{-1 + \bar{s}_l + \bar{s}_k} \, \mathcal{A}_n(z_j,\bar{z}_j)
\end{align}
Under the global conformal transformation (\ref{globalcon}), the transformed celestial amplitudes $\tilde{\mathcal{A}}_n $ transform as 
\begin{align}
    \tilde{\mathcal{A}}_n \to \prod_{j=1}^n(cz_j+d)^{-3+n-ns_j+2s_j-\sum_{j=1}^{n}s_j+\Delta_j+J_j} \nonumber \\(\bar{c}\bar{z}_j+\bar{d})^{-3+n-n\bar{s}_j+2\bar{s}_j-\sum_{j=1}^{n}\bar{s}_j+\Delta_j-J_j}\tilde{\mathcal{A}}_n 
\end{align}
Conformal invariance then requires
\begin{align}
    -3+n-ns_j+2s_j-\sum_{j=1}^{n}s_j+\Delta_j+J_j = 0 \\
    -3+n-n\bar{s}_j+2\bar{s}_j-\sum_{j=1}^{n}\bar{s}_j+\Delta_j-J_j = 0
\end{align}
We can express $(s_j,\bar{s}_j)$ in terms of $(\Delta_j,J_j)$ as follows: 
\begin{align}
    s_j = \frac{\Delta_j+J_j-3+n -\mathcal{S}}{n-2} \label{sj} \\
    \bar{s}_j =\frac{\Delta_j-J_j-3+n -\bar{\mathcal{S}}}{n-2} \label{barsj}
\end{align}
where $\mathcal{S} \equiv \sum_{j=1}^n s_j$ and $\bar{\mathcal{S}}\equiv \sum_{j=1}^n\bar{s}_j$. 
Conversely, the inverse relations can be written as
\begin{align}
    \Delta_j &= \frac{1}{2}\left((n-2)(s_j+\bar{s}_j)+6-2n +\mathcal{S}+\bar{\mathcal{S}}\right) \label{deltas}\\
    J_j &= \frac{1}{2}\left((n-2)(s_j-\bar{s}_j)+\mathcal{S}-\bar{\mathcal{S}}\right)\label{js}
\end{align}
Under the integral tranformation, the physical data of the external particles, $(\Delta_j,J_j)$, are encoded in the new variables $(s_j,\bar{s}_j)$.  Furthermore, by summing over $j$ from 1 to $n$ in (\ref{sj}) and (\ref{barsj}), we obtain 
\begin{align}
    \sum_{j=1}^n s_j &= \frac{n^2-3n+\sum_{j=1}^n(\Delta_j+J_j)}{2(n-1)}\\
    \sum_{j=1}^n\bar{s}_j &= \frac{n^2-3n+\sum_{j=1}^n(\Delta_j-J_j)}{2(n-1)}
\end{align}
These expressions are generalized versions of the conditions on the variables $(s_j,\bar{s}_j)$. Upon identifying the values of $(\Delta_j,J_j)$ for the MHV configurations, we recover the conditions given in (\ref{condNpoint}) and (\ref{condNgrav}).
\section{Conclusions}\label{conclusion}
In this work, we constructed an integral transformation for celestial gluon amplitudes, given in (\ref{trans1}), which maps the celestial complex coordinates $(z_i,\bar{z}_i)$ to a new set of complex variables $(s_i,\bar{s}_i)$. This transformation was designed to mimic the structure of closed string amplitudes, which are known to exhibit conformal invariance. For self-consistency, we also defined the corresponding inverse transformation. In constructing this inverse map, a potential divergence—reflecting the translational redundancy inherent in celestial amplitudes—was carefully identified and regulated. This divergence was subsequently absorbed into an overall normalization factor, leading to the well-behaved inverse transformation presented in \eqref{inverse}.

When performing the given integral transformation on the celestial MHV gluon amplitudes, we derived the constraints on the new variables $(s_i,\bar{s}_i)$. These constraints are given in (\ref{cond3point}), (\ref{cond4point}), and (\ref{condNpoint}) for the three-, four-, and general \(n\)-point cases respectively. Similarly, the constraints in (\ref{3condg}), (\ref{4condg}), and (\ref{condNgrav}) correspond to the three-, four-, and general \(n\)-point cases of celestial MHV graviton amplitudes, respectively.
We demonstrated that these conditions are necessary for the transformed amplitudes to remain invariant under global conformal transformations. 
Furthermore, the new variables $(s_j\bar{s}_j)$ are related to the physical data of the external particles through the relations given in (\ref{sj}) and (\ref{barsj}), while the inverse relations are given in (\ref{deltas}) and (\ref{js}).

Since the resulting conformally invariant celestial amplitudes are structurally reminiscent of closed string amplitudes, this allows future exploration on implementing techniques from string theory—such as holomorphic factorization—can be applied to analyze these new celestial amplitudes. However, such an extension is not entirely straightforward: the holomorphic and anti-holomorphic coordinates $z_i,\bar{z}_i$ are now intertwined through momentum conservation constraints which may lead to structural features that differ from those of conventional string amplitudes. Overall, the construction of these new conformally invariant celestial amplitudes provides a new perspective on the structure of celestial holography and motivates further investigations.

\section*{Acknowledgements}
AY is thankful for the support from Petchra Pra Jom Klao Ph.D. Scholarship from King Mongkut's University of Technology Thonburi. PS is grateful to the National Science, Research and Innovation Fund (NSRF) via the Program Management Unit for Human Resources \& Institutional Development, Research and Innovation for support under grant number B39G690007.

\appendix

\section{Dirac delta}
\label{nDelta}
The $n$-point delta function of momentum conservation can be written as \cite{npointDelta}
\begin{align}
    \delta^{(4)}(\sum_{j=1}^n \epsilon_j\omega_jq_j) = \frac{i}{4}\frac{(1-t_4)(1-\bar{t}_4)}{t_4-\bar{t}_4}\frac{1}{|z_{14}|^2|z_{23}|^2}\prod_{j=1}^{4}\delta(\omega_i-\omega_i^*)
\end{align}
where $\omega^*_i$ = $f_{i5}\omega_5+f_{i6}\omega_6+\dots+f_{in}\omega_n$ and $t_i$ is defined as 
\begin{align}
    t_j = \frac{z_{12}z_{3j}}{z_{13}z_{2j}}
\end{align}
The functions $f_{ij}$ are given by 
\begin{align}
    f_{1j} &= t_4 \left|\frac{z_{24}}{z_{12}}\right|^2 \frac{(1-t_4)(1-\bar{t}_4)}{t_4-\bar{t}_4}\epsilon_1\epsilon_j \frac{t_j-\bar{t}_j}{(1-t_j)(1-\bar{t}_j)}\left|\frac{z_{1j}}{z_{14}}\right|^2 \nonumber \\
    &\hspace{0.3cm}-\epsilon_1\epsilon_j t_j \left|\frac{z_{2j}}{z_{12}}\right|^2 
     \\
    f_{2j} &= -\frac{1-t_4}{t_4}\left|\frac{z_{34}}{z_{23}}\right|^2\frac{(1-t_4)(1-\bar{t}_4)}{t_4-\bar{t}_4}\frac{\epsilon_1\epsilon_j}{\epsilon_1\epsilon_2}\frac{t_j-\bar{t}_j}{(1-t_j)(1-\bar{t}_j)}\left|\frac{z_{1j}}{z_{14}}\right|^2 \nonumber \\
    &\hspace{0.3cm}+\frac{\epsilon_1\epsilon_j}{\epsilon_1\epsilon_2}\frac{1-t_j}{t_j}\left|\frac{z_{3j}}{z_{23}}\right|^2 \\
    f_{3j} &= (1-t_4)\left|\frac{z_{24}}{z_{23}}\right|^2\frac{(1-t_4)(1-\bar{t}_4)}{t_4-\bar{t}_4}\frac{\epsilon_1\epsilon_j}{\epsilon_1\epsilon_3}\frac{t_j-\bar{t}_j}{(1-t_j)(1-\bar{t}_j)}\left|\frac{z_{1j}}{z_{14}}\right|^2 \nonumber \\
    &\hspace{0.3cm}+\frac{\epsilon_1\epsilon_j}{\epsilon_1\epsilon_3}(1-t_j)\left|\frac{z_{2j}}{z_{23}}\right|^2 \\
    f_{4j}&= -\frac{(1-t_4)(1-\bar{t}_4)}{t_4-\bar{t}_4} \frac{\epsilon_1\epsilon_j}{\epsilon_1\epsilon_4}\frac{t_j-\bar{t}_j}{(1-t_j)(1-\bar{t}_j)}\left|\frac{z_{1j}}{z_{14}}\right|^2
\end{align}
where $j$ = $5,6,\dots, n$. This expression of delta function can be used for $n \geq 5$.
\section{Fractional derivative} 
\label{Fractional}
As we mentioned in section \ref{TestFor}, the fractional derivative was used. The fractional derivative is the generalization of the derivative with non-integer order. The fractional derivative is defined as 
\begin{align}
    _aD^{\alpha}_{x} f(x) \equiv \frac{1}{\Gamma(m-\alpha)}\frac{d^m}{dx^m} \int_{a}^xdk \,\, (x-k)^{m-\alpha-1}f(k)
 \end{align}
 where $\alpha \in \mathbb{C}$  with $\text{Re}(\alpha)>0$, $m \in \mathbb{N}$ with $m>\text{Re}(\alpha)$ and $a$ is the integration limit. The fractional derivative depends on the limit $a$ \cite{Reviewfractional1,miller1993introduction}. For the case 
\begin{align}
    f(x) = e^{\beta x}
\end{align}
The fractional derivative of $f(x)$ is given by \cite{miller1993introduction}
\begin{align}
    _{-\infty}D^{\alpha}_{x} e^{\beta x} =  \beta^{\alpha}e^{\beta x}
\end{align}
for arbitrary complex number $\alpha$ and $\text{Re}(\beta)>0$. Therefore, we can use above result to compute the derivative in (\ref{funcderiv}). We get 
\begin{align}
    I_1 = 2\pi i \Gamma{(n+1)}\int_{0}^{\infty}dt \,t^{-n}
\end{align}
for $n$ is non-integer with $\text{Re}(n)>0$.
\bibliographystyle{ieeetr} 
\bibliography{Ref}

@book{Elvang_Huang_2015, 
place={Cambridge}, 
title={Scattering Amplitudes in Gauge Theory and Gravity}, 
publisher={Cambridge University Press}, 
author={Elvang, Henriette and Huang, Yu-tin}, 
year={2015}}

@article{Puhm:2019zbl,
    author = "Puhm, Andrea",
    title = "{Conformally Soft Theorem in Gravity}",
    eprint = "1905.09799",
    archivePrefix = "arXiv",
    primaryClass = "hep-th",
    reportNumber = "CPHT-RR021.052019",
    doi = "10.1007/JHEP09(2020)130",
    journal = "JHEP",
    volume = "09",
    pages = "130",
    year = "2020"
}

@article{StringOnCS,
    author = "Stieberger, Stephan and Taylor, Tomasz R.",
    title = "{Strings on Celestial Sphere}",
    eprint = "1806.05688",
    archivePrefix = "arXiv",
    primaryClass = "hep-th",
    reportNumber = "MPP-2018-136",
    doi = "10.1016/j.nuclphysb.2018.08.019",
    journal = "Nucl. Phys. B",
    volume = "935",
    pages = "388--411",
    year = "2018"
}

@article{ParkeTaylor,
    author = "Parke, Stephen J. and Taylor, T. R.",
    title = "{An Amplitude for $n$ Gluon Scattering}",
    reportNumber = "FERMILAB-PUB-86-042-T",
    doi = "10.1103/PhysRevLett.56.2459",
    journal = "Phys. Rev. Lett.",
    volume = "56",
    pages = "2459",
    year = "1986"
}

@article{npointDelta,
    author = "Fan, Wei and Fotopoulos, Angelos and Taylor, Tomasz R.",
    title = "{Soft Limits of Yang-Mills Amplitudes and Conformal Correlators}",
    eprint = "1903.01676",
    archivePrefix = "arXiv",
    primaryClass = "hep-th",
    doi = "10.1007/JHEP05(2019)121",
    journal = "JHEP",
    volume = "05",
    pages = "121",
    year = "2019"
}

@article{Pasterski:2017ylz,
    author = "Pasterski, Sabrina and Shao, Shu-Heng and Strominger, Andrew",
    title = "{Gluon Amplitudes as 2d Conformal Correlators}",
    eprint = "1706.03917",
    archivePrefix = "arXiv",
    primaryClass = "hep-th",
    doi = "10.1103/PhysRevD.96.085006",
    journal = "Phys. Rev. D",
    volume = "96",
    number = "8",
    pages = "085006",
    year = "2017"
}

@book{Green_Schwarz_Witten_2012, place={Cambridge}, series={Cambridge Monographs on Mathematical Physics}, title={Superstring Theory: 25th Anniversary Edition}, publisher={Cambridge University Press}, author={Green, Michael B. and Schwarz, John H. and Witten, Edward}, year={2012}, collection={Cambridge Monographs on Mathematical Physics}}

@article{Reviewfractional1,
author = {de Oliveira, Edmundo Capelas and Tenreiro Machado, José António},
title = {A Review of Definitions for Fractional Derivatives and Integral},
journal = {Mathematical Problems in Engineering},
volume = {2014},
number = {1},
pages = {238459},
doi = {https://doi.org/10.1155/2014/238459},
url = {https://onlinelibrary.wiley.com/doi/abs/10.1155/2014/238459},
eprint = {https://onlinelibrary.wiley.com/doi/pdf/10.1155/2014/238459},
year = {2014}
}

@book{miller1993introduction,
  title={An Introduction to the Fractional Calculus and Fractional Differential Equations},
  author={Miller, K.S. and Ross, B.},
  isbn={9780471588849},
  lccn={lc93009500},
  url={https://books.google.co.th/books?id=MOp_QgAACAAJ},
  year={1993},
  publisher={Wiley}
}

@article{Pasterski:2016qvg,
    author = "Pasterski, Sabrina and Shao, Shu-Heng and Strominger, Andrew",
    title = "{Flat Space Amplitudes and Conformal Symmetry of the Celestial Sphere}",
    eprint = "1701.00049",
    archivePrefix = "arXiv",
    primaryClass = "hep-th",
    doi = "10.1103/PhysRevD.96.065026",
    journal = "Phys. Rev. D",
    volume = "96",
    number = "6",
    pages = "065026",
    year = "2017"
}

@article{Pasterski:2017kqt,
    author = "Pasterski, Sabrina and Shao, Shu-Heng",
    title = "{Conformal basis for flat space amplitudes}",
    eprint = "1705.01027",
    archivePrefix = "arXiv",
    primaryClass = "hep-th",
    doi = "10.1103/PhysRevD.96.065022",
    journal = "Phys. Rev. D",
    volume = "96",
    number = "6",
    pages = "065022",
    year = "2017"
}

@book{Strominger:2017zoo,
    author = "Strominger, Andrew",
    title = "{Lectures on the Infrared Structure of Gravity and Gauge Theory}",
    eprint = "1703.05448",
    archivePrefix = "arXiv",
    primaryClass = "hep-th",
    isbn = "978-0-691-17973-5",
    publisher = "Princeton University Press",
    year = "2018"
}

@misc{raclariu2021lecturescelestialholography,
      title={Lectures on Celestial Holography}, 
      author={Ana-Maria Raclariu},
      year={2021},
      eprint={2107.02075},
      archivePrefix={arXiv},
      primaryClass={hep-th},
      url={https://arxiv.org/abs/2107.02075}, 
}

@article{Yuenyong:2024ebe,
    author = "Yuenyong, Aphiwat and Srisangyingcharoen, Pongwit",
    title = "{Relations between closed string amplitudes and mixed string amplitudes at tree-level}",
    eprint = "2402.05775",
    archivePrefix = "arXiv",
    primaryClass = "hep-th",
    doi = "10.1007/JHEP08(2024)097",
    journal = "JHEP",
    volume = "08",
    pages = "097",
    year = "2024"
}

@article{Kawai:1985xq,
    author = "Kawai, H. and Lewellen, D. C. and Tye, S. H. H.",
    title = "{A Relation Between Tree Amplitudes of Closed and Open Strings}",
    reportNumber = "CLNS-85/667",
    doi = "10.1016/0550-3213(86)90362-7",
    journal = "Nucl. Phys. B",
    volume = "269",
    pages = "1--23",
    year = "1986"
}

@article{Hannah01051996,
author = {John Hannah},
title = {A Geometric Approach to Determinants},
journal = {The American Mathematical Monthly},
volume = {103},
number = {5},
pages = {401--409},
year = {1996},
publisher = {Taylor \& Francis},
doi = {10.1080/00029890.1996.12004759},
URL = { https://doi.org/10.1080/00029890.1996.12004759},
eprint = {  https://doi.org/10.1080/00029890.1996.12004759}
}

@article{bjerrum2009minimal,
  title={Minimal basis for gauge theory amplitudes},
  author={Bjerrum-Bohr, N Emil J and Damgaard, Poul H and Vanhove, Pierre},
  journal={Physical review letters},
  volume={103},
  number={16},
  pages={161602},
  year={2009},
  publisher={APS}
}

@article{SCHREIBER2018349,
title = {Tree-level gluon amplitudes on the celestial sphere},
author = {Anders {\o}. Schreiber and Anastasia Volovich and Michael Zlotnikov},
journal = {Physics Letters B},
volume = {781},
pages = {349-357},
year = {2018},
issn = {0370-2693},
doi = {https://doi.org/10.1016/j.physletb.2018.04.010},
url = {https://www.sciencedirect.com/science/article/pii/S0370269318302946},

}

@article{Donnay:2023kvm,
    author = "Donnay, Laura and Giribet, Gaston and Gonz{\'a}lez, Hern{\'a}n and Puhm, Andrea and Rojas, Francisco",
    title = "{Celestial open strings at one-loop}",
    eprint = "2307.03551",
    archivePrefix = "arXiv",
    primaryClass = "hep-th",
    reportNumber = "CPHT-RR036.072023",
    doi = "10.1007/JHEP10(2023)047",
    journal = "JHEP",
    volume = "10",
    pages = "047",
    year = "2023"
}

@article{CanazasGaray:2025xlh,
    author = "Canazas Garay, Anthonny F. and Giribet, Gaston and Parra-Cisterna, Yoel and Rojas, Francisco",
    title = "{Celestial closed strings at one loop}",
    eprint = "2504.17989",
    archivePrefix = "arXiv",
    primaryClass = "hep-th",
    doi = "10.1103/3jqt-k1sl",
    journal = "Phys. Rev. D",
    volume = "111",
    number = "12",
    pages = "126014",
    year = "2025"
}

@article{Bern:2002kj,
    author = "Bern, Zvi",
    title = "{Perturbative quantum gravity and its relation to gauge theory}",
    eprint = "gr-qc/0206071",
    archivePrefix = "arXiv",
    reportNumber = "UCLA-02-TEP-9",
    doi = "10.12942/lrr-2002-5",
    journal = "Living Rev. Rel.",
    volume = "5",
    pages = "5",
    year = "2002"
}

@article{Bern:2008qj,
    author = "Bern, Z. and Carrasco, J. J. M. and Johansson, Henrik",
    title = "{New Relations for Gauge-Theory Amplitudes}",
    eprint = "0805.3993",
    archivePrefix = "arXiv",
    primaryClass = "hep-ph",
    reportNumber = "UCLA-07-TEP-15",
    doi = "10.1103/PhysRevD.78.085011",
    journal = "Phys. Rev. D",
    volume = "78",
    pages = "085011",
    year = "2008"
}

@article{Bern:2010ue,
    author = "Bern, Zvi and Carrasco, John Joseph M. and Johansson, Henrik",
    title = "{Perturbative Quantum Gravity as a Double Copy of Gauge Theory}",
    eprint = "1004.0476",
    archivePrefix = "arXiv",
    primaryClass = "hep-th",
    reportNumber = "UCLA-10-TEP-102, SACLAY-IPHT-T10-044",
    doi = "10.1103/PhysRevLett.105.061602",
    journal = "Phys. Rev. Lett.",
    volume = "105",
    pages = "061602",
    year = "2010"
}

@article{Bern:2010yg,
    author = "Bern, Zvi and Dennen, Tristan and Huang, Yu-tin and Kiermaier, Michael",
    title = "{Gravity as the Square of Gauge Theory}",
    eprint = "1004.0693",
    archivePrefix = "arXiv",
    primaryClass = "hep-th",
    reportNumber = "UCLA-TEP-10-103, PUPT-2335",
    doi = "10.1103/PhysRevD.82.065003",
    journal = "Phys. Rev. D",
    volume = "82",
    pages = "065003",
    year = "2010"
}

@article{Casali:2020vuy,
    author = "Casali, Eduardo and Puhm, Andrea",
    title = "{Double Copy for Celestial Amplitudes}",
    eprint = "2007.15027",
    archivePrefix = "arXiv",
    primaryClass = "hep-th",
    reportNumber = "CPHT-RR049.072020",
    doi = "10.1103/PhysRevLett.126.101602",
    journal = "Phys. Rev. Lett.",
    volume = "126",
    number = "10",
    pages = "101602",
    year = "2021"
}

@article{deBoer:2003vf,
    author = "de Boer, Jan and Solodukhin, Sergey N.",
    title = "{A Holographic reduction of Minkowski space-time}",
    eprint = "hep-th/0303006",
    archivePrefix = "arXiv",
    reportNumber = "ITFA-2003-11",
    doi = "10.1016/S0550-3213(03)00494-2",
    journal = "Nucl. Phys. B",
    volume = "665",
    pages = "545--593",
    year = "2003"
}

@article{Strominger:2013lka,
    author = "Strominger, Andrew",
    title = "{Asymptotic Symmetries of Yang-Mills Theory}",
    eprint = "1308.0589",
    archivePrefix = "arXiv",
    primaryClass = "hep-th",
    doi = "10.1007/JHEP07(2014)151",
    journal = "JHEP",
    volume = "07",
    pages = "151",
    year = "2014"
}

@article{Strominger:2013jfa,
    author = "Strominger, Andrew",
    title = "{On BMS Invariance of Gravitational Scattering}",
    eprint = "1312.2229",
    archivePrefix = "arXiv",
    primaryClass = "hep-th",
    doi = "10.1007/JHEP07(2014)152",
    journal = "JHEP",
    volume = "07",
    pages = "152",
    year = "2014"
}

@article{Stieberger:2009hq,
    author = "Stieberger, S.",
    title = "{Open {\&} Closed vs. Pure Open String Disk Amplitudes}",
    eprint = "0907.2211",
    archivePrefix = "arXiv",
    primaryClass = "hep-th",
    reportNumber = "MPP-2008-01",
    month = "7",
    year = "2009"
}

@article{Stieberger:2015vya,
    author = "Stieberger, Stephan and Taylor, Tomasz R.",
    title = "{Disk Scattering of Open and Closed Strings (I)}",
    eprint = "1510.01774",
    archivePrefix = "arXiv",
    primaryClass = "hep-th",
    reportNumber = "MPP-2015-184, MPP--2015--184",
    doi = "10.1016/j.nuclphysb.2015.12.002",
    journal = "Nucl. Phys. B",
    volume = "903",
    pages = "104--117",
    year = "2016"
}

@article{Boels:2010bv,
    author = "Boels, Rutger H. and Marmiroli, Daniele and Obers, Niels A.",
    title = "{On-shell Recursion in String Theory}",
    eprint = "1002.5029",
    archivePrefix = "arXiv",
    primaryClass = "hep-th",
    doi = "10.1007/JHEP10(2010)034",
    journal = "JHEP",
    volume = "10",
    pages = "034",
    year = "2010"
}

@article{Chang:2012qs,
    author = "Chang, Yung-Yeh and Feng, Bo and Fu, Chih-Hao and Lee, Jen-Chi and Wang, Yihong and Yang, Yi",
    title = "{A note on on-shell recursion relation of string amplitudes}",
    eprint = "1210.1776",
    archivePrefix = "arXiv",
    primaryClass = "hep-th",
    doi = "10.1007/JHEP02(2013)028",
    journal = "JHEP",
    volume = "02",
    pages = "028",
    year = "2013"
}

@article{Srisangyingcharoen:2024qyx,
    author = "Srisangyingcharoen, Pongwit",
    title = "{General expressions for on-shell recursion relations for tree-level open string amplitudes}",
    eprint = "2404.00244",
    archivePrefix = "arXiv",
    primaryClass = "hep-th",
    doi = "10.1016/j.physletb.2024.139038",
    journal = "Phys. Lett. B",
    volume = "858",
    pages = "139038",
    year = "2024"
}

@article{Srisangyingcharoen:2024zko,
    author = "Srisangyingcharoen, Pongwit and Yuenyong, Aphiwat",
    title = "{On-shell recursion relations for tree-level closed string amplitudes}",
    eprint = "2410.15448",
    archivePrefix = "arXiv",
    primaryClass = "hep-th",
    doi = "10.1140/epjc/s10052-025-14858-8",
    journal = "Eur. Phys. J. C",
    volume = "85",
    number = "10",
    pages = "1118",
    year = "2025"
}

@article{Adamo:2024mqn,
    author = "Adamo, Tim and Bu, Wei and Tourkine, Piotr and Zhu, Bin",
    title = "{Eikonal amplitudes on the celestial sphere}",
    eprint = "2405.15594",
    archivePrefix = "arXiv",
    primaryClass = "hep-th",
    doi = "10.1007/JHEP10(2024)192",
    journal = "JHEP",
    volume = "10",
    pages = "192",
    year = "2024"
}



%% For example:

%% \bibitem[Aladro et al.(2015)]{Aladro15} Aladro, R., Martín, S., Riquelme, D., et al. 2015, \aas, 579, A101

%%\end{thebibliography}

\end{document}